\documentclass[12pt,preprint]{aastex}

\usepackage{xcolor}

\newcommand{\bdv}[1]{\mbox{\boldmath$#1$}}

\def\au{{\rm AU}} 
 
\def\kms{{\rm km}\,{\rm s}^{-1}}
\def\masyr{{\rm mas}\,{\rm yr}^{-1}}
\def\kpc{{\rm kpc}}
\def\mas{{\rm mas}}

\def\muas{\mu{\rm as}}

\def\max{{\rm max}}

\def\rel{{\rm rel}}

\def\eff{{\rm eff}}

\def\e{{\rm E}}
\def\bpi{{\bdv\pi}}
\def\bmu{{\bdv\mu}}

\begin{document}
\title{KMT-2019-BLG-0842Lb: A Cold Planet Below the Uranus/Sun Mass Ratio}

\input author.list 

\begin{abstract}
We report the discovery of a cold planet with a very low planet/host mass ratio of 
$q=(4.09\pm0.27) \times 10^{-5}$, which is similar to the ratio of Uranus/Sun ($q=4.37 \times 10^{-5}$) 
in the Solar system. The Bayesian estimates for the host mass, planet mass, system distance, 
and planet-host projected separation are 
$M_{\rm host}=0.76\pm 0.40 M_\odot$, 
$M_{\rm planet}=10.3\pm 5.5 M_\oplus$, 
$D_{\rm L} = 3.3\pm 1.3\,\kpc$, and 
$a_\perp = 3.3\pm 1.4\,\au$, respectively. 
The consistency of the color and brightness expected from the estimated lens mass and distance 
with those of the blend suggests the possibility that the most blended light comes from the planet 
host, and this hypothesis can be established if high resolution images are taken during the next 
(2020) bulge season. We discuss the importance of conducting optimized photometry and aggressive 
follow-up observations for moderately or very high magnification events 
to maximize the detection rate of planets with very low mass ratios.
\end{abstract}

\section{{Introduction}
\label{sec:intro}}

For the first 15 years of 
microlensing planet detections, there was no clear evidence for
``cold planets'' (i.e., beyond the snow line) with mass ratios below 
that of Uranus $(q=4.37\times 10^{-5})$ neither in our Solar System nor 
in other systems.  There are of course several cold, low-mass bodies
in the outer Solar System, but these are more than 10 times
less massive than the smallest terrestrial planet (Mercury).  The same
applies to Ceres, which lies approximately on the snow line.  Such
bodies likely form by a distinct mechanism relative to planets, and
they were therefore reclassified as ``dwarf planets'' by the IAU.

Cold, low-mass planets in exo-systems can only be detected by
microlensing \citep{bennett96,ob05390,ob05169}. However, there was 
no discovery of such a planet until one with a mass ratio $q=4.7\times 10^{-5}$ 
was reported in 2017 \citep{ob171434}\footnote{Within their sample of seven 
$q<1\times 10^{-4}$ microlensing planets, OGLE-2016-BLG-1195, had the lowest 
mass mass ratio $q=(4.7\pm 0.5)\times 10^{-5}$. 
\citet{ob171434} took the average from two groups who had made independent 
parameter determinations using disjoint data sets:
$q=[(4.2 \pm 0.7) \& (5.5 \pm 0.8)]\times 10^{-5}$ from \citet{ob161195a} and 
\citet{ob161195b}, respectively. \citet{ob171434} excluded OGLE-2017-BLG-0173
from their sample because its solution was ambiguous by a factor 2.5
($2.48\pm 0.24$ versus $6.4\pm 1.0$)$\times 10^{-5}$ at $\Delta\chi^2=3.5$
\citep{ob170173}.}.

In principle, the lack of such detections could have been due to one of the 
following four causes: (1) poor sensitivity of microlensing searches to such
planets; (2) extreme rarity (or complete absence) of such planets in
nature; (3) adverse fluctuation of small-number statistics; or (4)
some combination of these.  \citet{suzuki16} showed that there was a break in the
power-law distribution (in $q$) below some threshold, estimated
roughly at $q_{\rm br}\sim 1.7\times 10^{-4}$, arguing in particular
that their MOA-based planet sample had significant sensitivity below
the lowest-$q$ detections.  \citet{ob171434} confirmed that 
if the mass ratio distribution of planets with $q<1\times 10^{-4}$ 
were modeled as a power law, its slope was rising with $q$, 
in contrast to the well-established falling power law
at higher $q$.  Moreover, they showed that four of the
seven planets that they analyzed would have been detected even if their
mass ratios had been below $q=3\times 10^{-5}$, and one of these would
have been detected even at $q=2\times 10^{-6}$, i.e., below the Earth/Sun
mass ratio.  \citet{kb170165} argued that the dearth of planets
below $q=4.7\times 10^{-5}$ was due either to a break in the mass-ratio
function at $q_{\rm br}\simeq 5.6\times 10^{-5}$ or a ``pile-up'' of
planets near the Neptune-mass-ratio threshold.

With a mass ratio $q=(1.8\pm 0.2)\times 10^{-5}$, KMT-2018-BLG-0029
was the first reported planet that lay clearly below the previous
apparent Uranus/Sun mass-ratio threshold \citep{kb180029}.  It 
provided proof for the existence of such planets, but a single
detection yielded very limited information on their frequency.
\citet{kb180029} reviewed the history of the (by then, nine)
$q<1\times 10^{-4}$ planet detections, and concluded that with
the advent of regular observations of the
Korea Microlensing Telescope Network (KMTNet, 
\citealt{eventfinder,2016k2,2016eventfinder}) in 2016,
the rate of such detections would be increasing, and that it should
be possible to probe the frequency of such low-mass planets within a few
years.


Although microlensing is sensitive to planets with mass ratios below $q\sim10^{-4}$, 
the number of known microlensing planets with very low mass ratios 
is still small and these planets comprise a very minor fraction 
of all microlensing planets. Therefore, detecting more planets 
in the low mass regime is important to investigate the physical 
parameter distributions of these planets and to compare the distributions 
with those of planets in the higher mass regime. In this paper, we report the
detection of a cold planet with a mass ratio of $q=(4.09\pm0.27) \times 10^{-5}$.

\section{{Observations}
\label{sec:obs}}

KMT-2019-BLG-0842, (R.A., Decl.)$_{\rm J2000}$=(17:53:50.03, $-29$:52:38.78),
corresponding to $(l,b)=(+0.11,-2.02)$,
was discovered by the KMT alert-finder system
\citep{alertfinder} and announced as a microlensing candidate 
on the KMTNet website\footnote{http://kmtnet.kasi.re.kr/$\sim$ulens} 
at UT 02:39 on 16 May 2019, about six days before the event reached its peak.
The event was independently found by the Optical 
Gravitational Lensing Experiment (OGLE, \citealt{ogle-iv})
and the Microlensing Observations in Astrophysics (MOA, \citealt{ob03235})
collaborations
at UT 19:30 on 18 May 2019 as OGLE-2019-BLG-0763 and 
at UT 07:02 on 24 May 2019 as MOA-2019-BLG-232, respectively.

Observations by the KMTNet survey were conducted utilizing three identical 
1.6m telescopes equipped with $(2^\circ\times 2^\circ)$ cameras, 
located at Cerro Tololo InterAmerican Observatory (KMTC), 
the South African Astronomical Observatory (KMTS), 
and the Siding Springs Observatory (KMTA) \citep{kmtnet}. 
KMT-2019-BLG-0842 lies in the overlapping KMT fields BLG02 and BLG42, 
each of which is observed at a nominal cadence of $\Gamma=2\,{\rm hr}^{-1}$. 
However, both fields were observed at an adjusted cadence of 
$\Gamma=3\,{\rm hr}^{-1}$ from KMTS and KMTA prior to 15 June, 
i.e., until more than three weeks after the peak of the event. 
Hence, the effective cadence was $\Gamma=4\,{\rm hr}^{-1}$ from KMTC and 
$\Gamma=6\,{\rm hr}^{-1}$ from KMTS and KMTA. The observations 
were primarily conducted in the $I$ band, but every tenth such 
observation was matched by a $V$-band observation in order to 
measure the source color.

The OGLE collaboration observed the event, located in their BLG501 field, 
with $\Gamma=1\,{\rm hr}^{-1}$ cadence using the 1.3m telescope, equipped with 
a $1.4\,{\rm deg}^2$ camera, located at Las Campanas Observatory in Chile. 
OGLE observations were also primarily conducted in the $I$ band.

\begin{figure}
\plotone{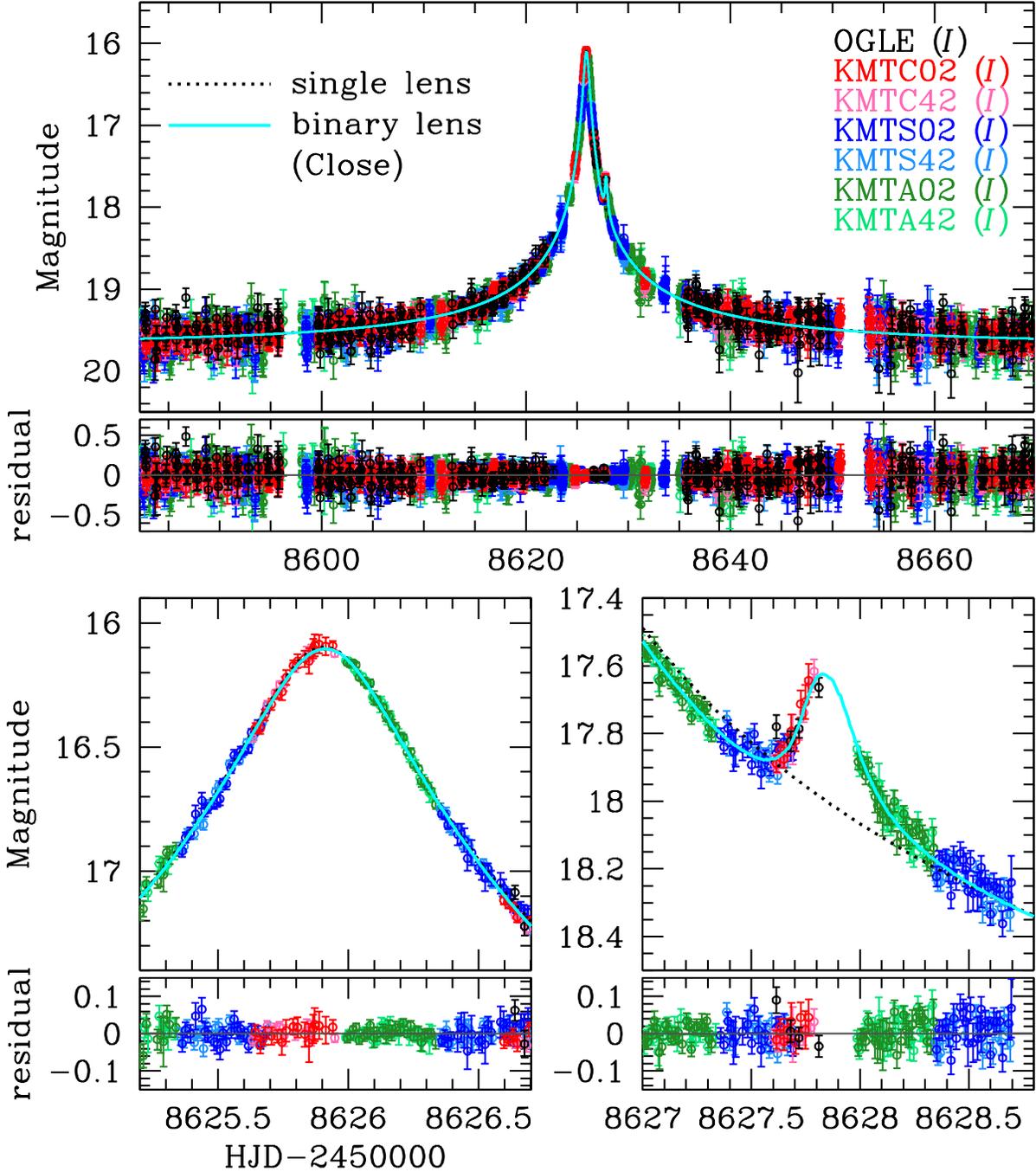}
\caption{Light curve and ``close'' 2L1S model for KMT-2019-BLG-0842.
Zooms of the peak and the anomaly are shown in the lower-left and lower-right
panels, respectively.
The anomaly lasts about 15 hours despite the very low planet-host
mass ratio $q=4.1\times 10^{-5}$ because the anomaly occurs
$\sim 7\times t_\eff$ effective timescales after peak, implying
that the trajectory is at an acute angle $\alpha\simeq \cot^{-1} 7\sim 8.5^\circ$
relative to the planet-host axis.
}
\label{fig:lc}
\end{figure}

The light curve derived from the KMTNet and OGLE observations is shown in 
Figure~\ref{fig:lc}.  The KMTNet data were reduced using pySIS \citep{albrow09},
which is a variant of difference image analysis (DIA, \citealt{tomaney96,alard98,alard00}). 
The OGLE data were reduced using another variant of DIA \citep{wozniak2000}. 
The relatively brief ($\sim 15\,$hr) planetary anomaly was announced by W.\ Zang 
at UT 19:56 on 26 May 2019 (JD$^\prime=$ JD-2450000 = 8630.33),
but by this time the anomaly was already over, and thus no followup observations
of the anomaly resulted from the announcement.

\begin{figure}
\plotone{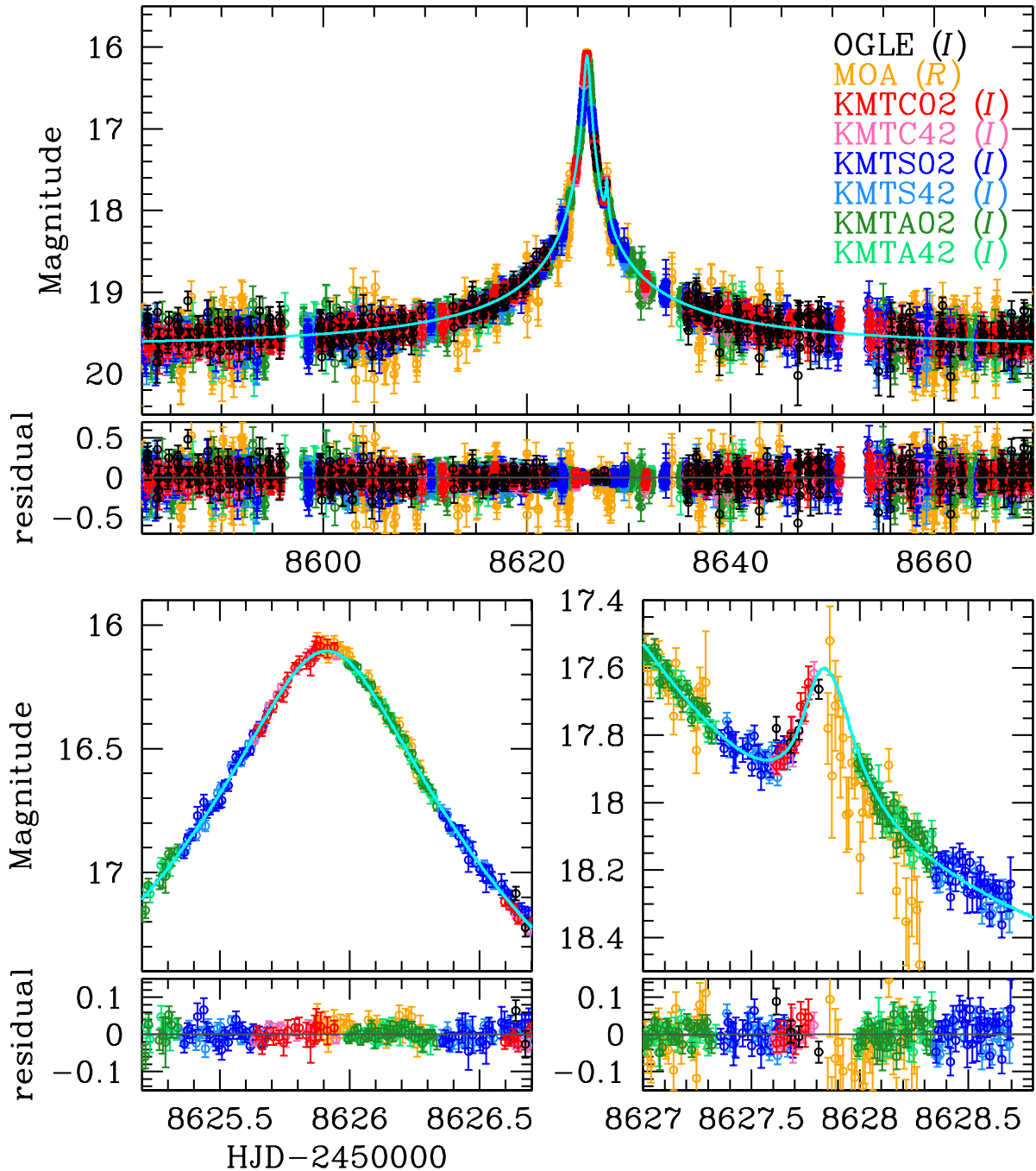}
\caption{Light curve KMT-2019-BLG-0842, as shown in Figure~\ref{fig:lc},
but with MOA data aligned to the best fit model.  The MOA data confirm
the anomaly but do not contribute to constraining the model and so
are not included in the fit.
}
\label{fig:lcmoa}
\end{figure}

The MOA collaboration observed the object with a cadence of $\Gamma=4\,{\rm hr}^{-1}$
using its 1.8m telescope at Mt.\ John, New Zealand, which is a equipped
with a $2.2\,{\rm deg^2}$ camera. However, observational conditions were 
poor on the night of the anomaly. The resulting observations therefore 
do not constrain the characteristics of the planet, although they do 
qualitatively confirm its existence. We therefore do not include these 
data in the primary fit, but rather use them to illustrate this confirmation (Figure~\ref{fig:lcmoa}). 
MOA observations were mainly taken using a broad $R/I$ filter.

\section{{Light Curve Analysis}
\label{sec:anal}}

With the exception of the short anomaly peaking at $t_{0,\rm anom}=8627.85$,
KMT-2019-BLG-0842 has the general appearance of a \citet{pac86} single-lens
single-source (1L1S) event. The light curve of a 1L1S event is characterized 
by three geometric parameters $(t_0,u_0,t_\e)$, respectively the time of maximum, the
impact parameter (in units of the Einstein radius $\theta_\e$), and
the Einstein radius crossing time,
\begin{equation}
t_\e\equiv {\theta_\e\over\mu_\rel};
\qquad
\theta_\e^2\equiv \kappa M\pi_\rel;
\qquad
\kappa\equiv {4G\over c^2\au}\simeq 8.14\,{\mas\over M_\odot},
\label{eqn:tedef}
\end{equation}
where $M$ is the lens mass and $(\pi_\rel,\bmu_\rel)$ are the lens-source
relative parallax and proper motion, respectively.  From visual inspection,
$t_0\simeq 8625.91$ and the full width at half maximum (FWHM) of the
light curve is 1.0 days. The peak magnification is either moderately high 
or very high depending on the blending, which is difficult to estimate by eye, 
and in either case $u_0\ll 1$, so $t_\eff\equiv u_0 t_\e = {\rm FWHM}/\sqrt{12}\simeq 0.29\,$day.
Hence (e.g., \citealt{gouldloeb}), the source almost certainly passes 
at an angle 
\begin{equation}
\alpha = \tan^{-1}{t_\eff \over t_{0,\rm anom} - t_0} = \tan^{-1}0.149 = 0.148,
\label{eqn:alphaeval}
\end{equation}
or $8.5^\circ$, relative to the binary axis.  The fact that the
anomaly is very short, even though the source crosses
the caustic structure at such an acute angle $\alpha$, 
suggests that the mass ratio between the binary lens components is very low. 
A similar inference can be drawn from 
the fact that there is no noticeable anomaly over the high 
(or very high) magnification peak.  That is, both effects tend to
indicate a small, resonant (or near-resonant) caustic.
However, detailed modeling is required to proceed further.

\subsection{{Binary Lens (2L1S) Analysis}
\label{sec:2L1S}}

We model the light curve as a 2L1S event with seven non-linear parameters 
$(t_0,u_0,t_\e,s,q,\alpha,\rho)$, where $s$ is the projected separation
between the binary components (normalized to $\theta_\e$),
$\rho\equiv \theta_*/\theta_\e$, and $\theta_*$ is the angular source
radius.  Notwithstanding the above estimate of $\alpha$, we conduct 
a broad grid search in the parameters $s$, $q$, and $\alpha$, 
in which $(s,q)$ are held
fixed and the remaining five parameters are allowed to vary in a 
Markov chain Monte Carlo (MCMC).  The three parameters $(t_0,u_0,t_\e)$
are seeded at their \citet{pac86} values, $\rho$ is seeded at $10^{-3}$,
and $\alpha$ is seeded at six values drawn uniformly from the unit
circle.  There are two flux parameters $(f_{s,i},f_{b,i})$
for each observatory $i$, which are fit by linear regression to 
the observed flux $F_i(t)$ for each model, 
according to $F_i(t) = f_{s,i}A(t) + f_{b,i}$.  These are identified with
the ``source'' and ``blend'' flux, respectively.

\begin{figure}
\plotone{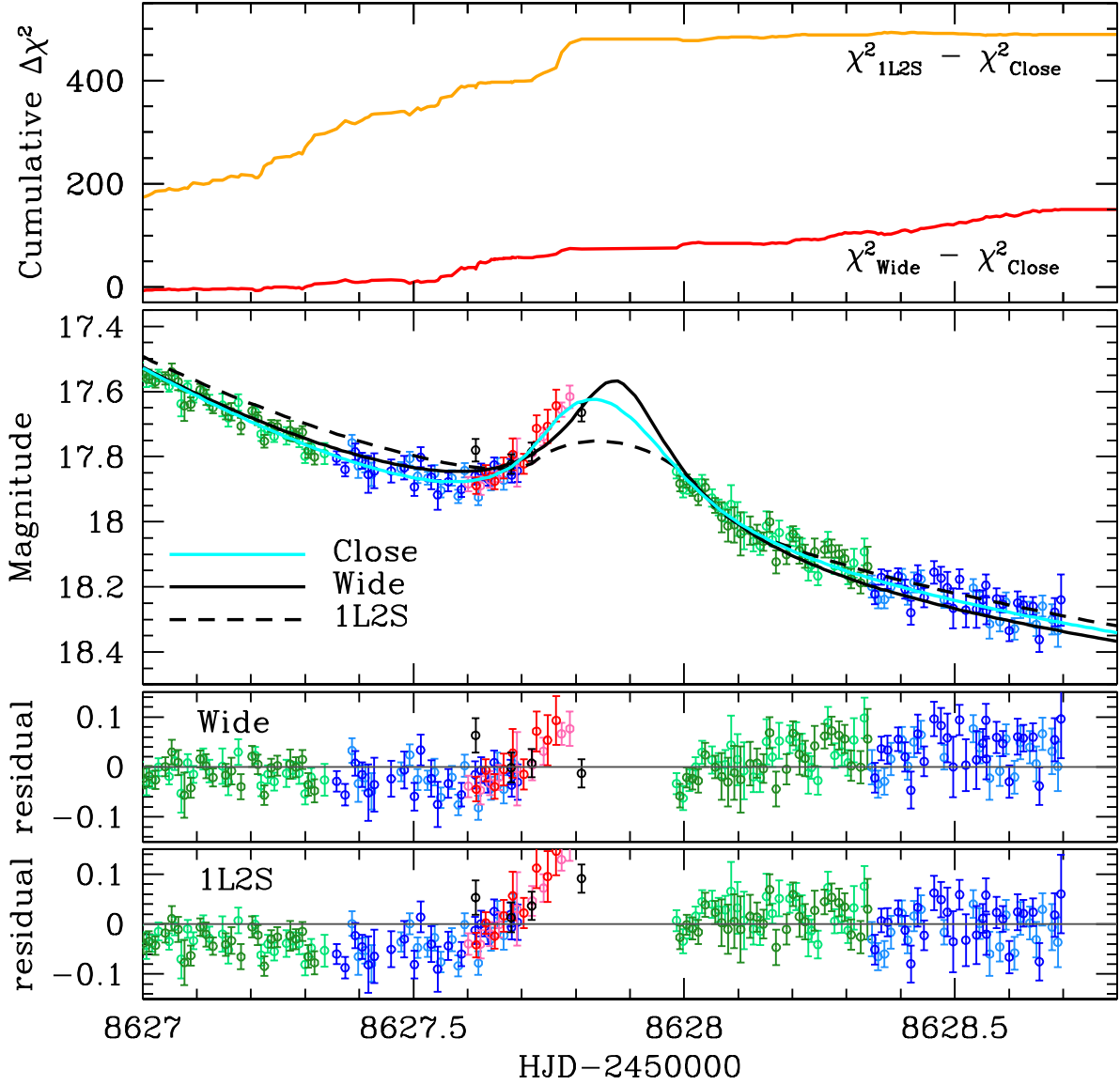}
\caption{Upper panel: $\Delta\chi^2$ difference 
relative to the only surviving model (2L1S ``close'') of 
two other possible models, 2L1S ``wide'' (red) and 1L2S (yellow)
for KMT-2019-BLG-0842.  The second panel shows the anomaly region
of the light curve together with these three models.  The lower
two panels show the residuals for the two excluded models.
See Figure~\ref{fig:lc} for the corresponding residuals of the
surviving (2L1S ``close'') model.
}
\label{fig:resid}
\end{figure}

\begin{figure}
\plotone{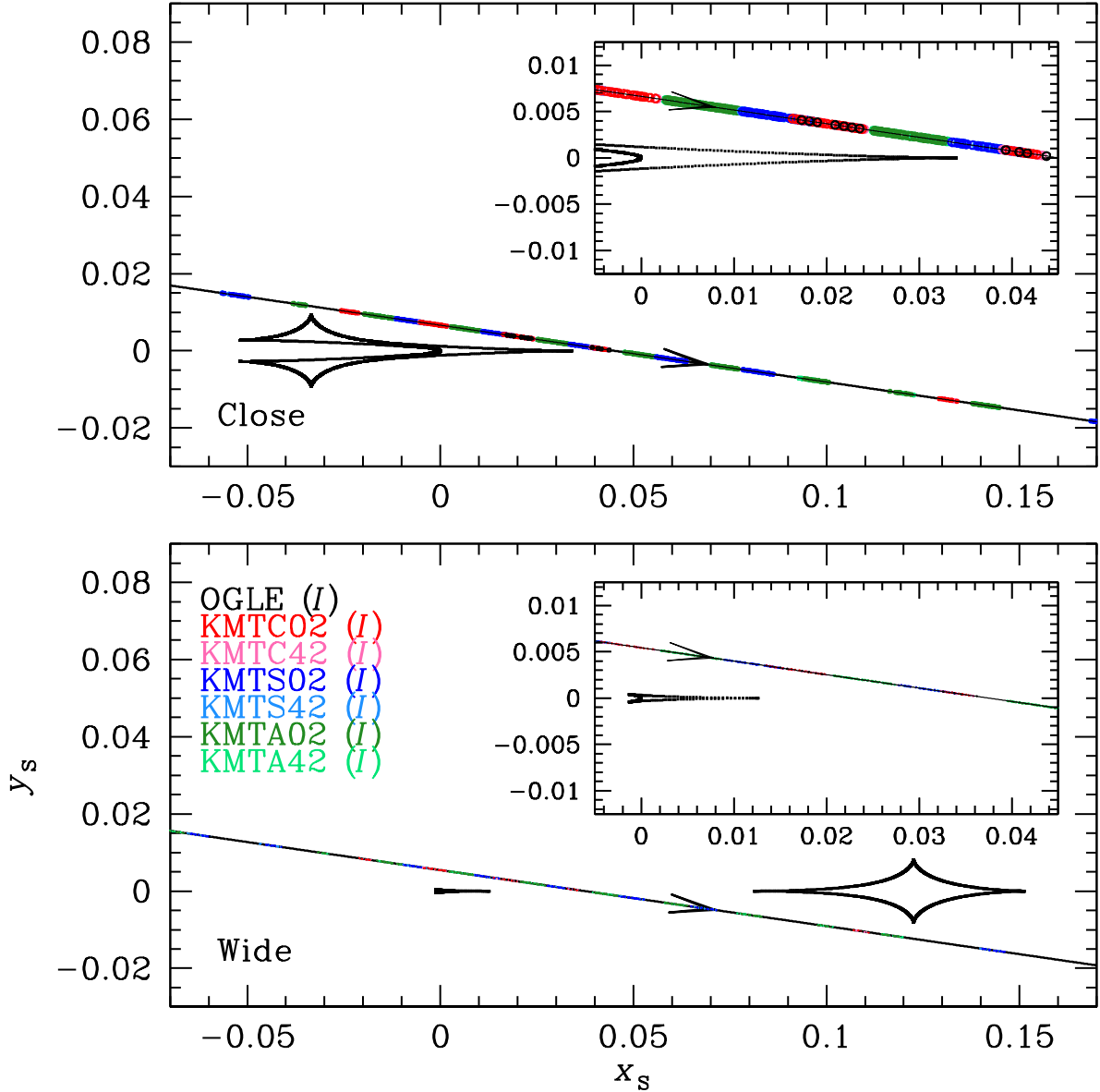}
\caption{Caustic geometries for the two possible 2L1S solutions
for KMT-2019-BLG-0842, i.e., ``close'' (top) and ``wide'' (bottom).
In both cases, the anomaly (see zooms in Figures~\ref{fig:lc} 
and \ref{fig:resid}) is generated when the source passes
over the planet-star axis about 1.94 days ($\simeq 0.044$ Einstein units)
after the peak.  In both cases, the anomaly is due to a narrow
high-magnification ridge that extends from the narrow end of
a caustic centered on the host.  For the ``close'' geometry, this
caustic has a resonant (six-sided) topology, while for the ``wide''
geometry it is a central caustic that is connected to the planetary
caustic by the ridge.  However, the ``wide'' geometry is excluded
by $\Delta\chi^2=176$.  See text and Figure~\ref{fig:resid}.
}
\label{fig:geo}
\end{figure}

We find only two local minima in the resulting $(s,q)$ map, which we then
further refine by allowing all seven parameters to vary in the MCMC.
Finally, these converge to $(s,q)=(0.98,4.1\times 10^{-5})$ and
$(s,q)=(1.06,3.9\times 10^{-5})$. See Table~\ref{tab:ulens}.
These solutions are actually quite similar in all of their parameters,
and in particular both agree with the analytically estimated value of 
the source trajectory angle $\alpha$ in Equation~(\ref{eqn:alphaeval}). 
However, the ``wide'' solution $(s>1)$ is disfavored relative
to the ``close'' solution $(s<1)$ by $\Delta\chi^2=176$.  Furthermore,
the ``wide'' solution has clear systematic residuals.  
See Figure~\ref{fig:resid}.  Hence, we exclude the ``wide'' solution.
The caustic geometries for both the ``close'' and ``wide'' solutions
are shown in Figure~\ref{fig:geo}.

We investigate whether the microlens parallax vector $\bpi_\e$ 
\citep{gould92,gould00,gould04} can be meaningfully constrained, but
we find that it cannot.

\subsection{{Binary Source (1L2S) Analysis}
\label{sec:1L2S}}

\citet{gaudi98} pointed out that a short-lived ``bump'' on an otherwise
normal \citet{pac86} light curve could in principle be produced by
a second source (1L2S) rather than a second lens (2L1S). This is of
particular concern when (as in the present case) 
the ``bump'' does not exhibit any obvious caustic structure. 
Therefore, we check the degeneracy between the 2L1S and 1L2S 
interpretations by additionally conducting a 1L2S modeling. 
Our 1L2S model has eight nonlinear parameters: $2\times(t_0,u_0,\rho)$
for the two sources, plus a common Einstein timescale $t_\e$ and
the $I$-band flux ratio between the two sources, $q_{F,I}$.
See Table~\ref{tab:1L2S}.

We find that the 2L1S interpretation is favored over the 1L2S interpretation by 
$\Delta\chi^2 = \chi^{2}({\rm 1L2S}) - \chi^{2}({\rm 2L1S}) = 495$. 
To check the region of the fit difference, we also inspect the cumulative $\Delta\chi^2$ 
distribution of $\Delta\chi^2 = \chi^{2}({\rm 1L2S}) - \chi^{2}({\rm 2L1S})$ (see Figure~\ref{fig:resid}). 
From this, we find that the $\chi^2$ difference largely comes from the anomaly region, 
in which the 1L2S solution provides a poorer fit to the observed light curve, 
especially in the rising part of the anomaly. 
In addition, the solution is unphysical in the sense that the flux ratio is 
$q_{F,I}\simeq 0.013$ while the normalized sources sizes $(\rho_1,\rho_2)$ 
are of the same order\footnote{Strictly speaking, $\rho_1$ is poorly constrained by the fit, 
but it has a strict upper limit $\rho_1<u_{0,1}=56\times 10^{-4}$, 
which implies the same unphysicality.}. See Table~\ref{tab:1L2S}. 
Hence, we exclude the 1L2S solution.


\section{{Color-Magnitude Diagram and Einstein Radius}
\label{sec:cmd}}

Although the planetary anomaly does not exhibit obvious caustic features,
the results in Table~\ref{tab:ulens} show that the normalized
source size $\rho$ is reasonably well measured\footnote{This was
also the case for the low-$q$ planetary event 
OGLE-2016-BLG-1195 \citep{ob161195a,ob161195b}}.
This implies that we can estimate $\theta_\e=\theta_*/\rho$ provided
that $\theta_*$ is measured. Indeed, although the $1\,\sigma$ error 
in $\rho$ is relatively large, there is a very strong, $3\,\sigma$, 
upper limit $\rho<5.7\times 10^{-4}$, which will place important 
physical constraints on the lens system, as we discuss below.

We follow the standard approach \citep{ob03262} of measuring $\theta_*$
by placing the source on an instrumental color-magnitude diagram (CMD).
See Figure~\ref{fig:cmd}.
For this purpose, we reduce the KMTC42 $V$ and $I$ data using pyDIA, 
which yields field-star and light-curve photometry on the same system.
Using this instrumental system, we find source and red-clump-centroid positions
of
$[(V-I),I]_{\rm S} = (2.38,21.73)\pm (0.03,0.01)$ and
$[(V-I),I]_{\rm cl} = (2.43,16.04)\pm (0.03,0.07)$, respectively, and hence
an offset of
$\Delta[(V-I),I] = (-0.05,5.69)\pm (0.04,0.07)$.
We adopt $[(V-I),I]_{\rm cl,0} = (1.06,14.44)$ \citep{bensby13,nataf13}.
We do not assign an error to this determination but rather add 5\% error
in quadrature to the final result for all aspects of the method. 
We then obtain $[(V-I),I]_{\rm S,0}=[(V-I),I]_{\rm cl,0} + \Delta[(V-I),I] = 
(1.01,20.13)\pm (0.04,0.07)$. Assuming that the source 
is located at the mean distance to the bulge, i.e., $D_{\rm S} = 8.17\,{\rm kpc}$ \citep{nataf13}, 
the source star has the color and absolute magnitude of $[(V-I)_0,M_I]_{\rm S}\simeq(1.0,5.6)$, 
which is typical of an early K dwarf.

We then convert the measured $V/I$ color into $V/K$ color 
using the $VIK$ color-color relations of \citet{bb88} and 
then apply the color/surface-brightness relations of \citet{kervella04} 
to obtain (after adding 5\% error),
\begin{equation}
\theta_* = 0.416\pm 0.031\ \muas.
\label{eqn:thetastar}
\end{equation}
We can then obtain ``naive'' estimates for $\theta_\e$ and $\mu_\rel$,
\begin{equation}
\theta_\e = {\theta_*\over\rho}= 0.96\pm 0.22\ \mas;
\qquad
\mu_\rel = 8.0\pm 1.8\ \masyr 
\qquad (\rm naive).
\label{eqn:thetaemu}
\end{equation}
We refer to these two estimates as ``naive'' because, as will be discussed 
in Section~\ref{sec:phys}, the proper evaluation of these quantities should 
be estimated from their Bayesian posteriors.

The estimated values of $\theta_\e$ and $\mu_\rel$
strongly imply that the lens lies in the disk. That is,
from the definition of $\theta_\e$ (Equation~(\ref{eqn:tedef})),
$\pi_\rel = 0.12\,\mas (\theta_\e/\mas)^2/(M/M_\odot)$. Thus, if we
adopt the best-fit value of $\theta_\e$ and also take account of the
fact that lenses
significantly more massive than the Sun would be easily visible,
then we would infer $\pi_\rel \ga 0.12\,\mas$, i.e., lens
distance $D_{\rm L}\la 4\,\kpc$.  Because the error in $\theta_\e$ is
large (due primarily to the large error in $\rho$), we must
also consider how smaller values of $\theta_\e$ (i.e., below the best-fit
value) would affect this
argument.  However, as mentioned above, the errors in $\rho$ are
highly asymmetric, so there is a very hard lower limit $\theta_\e > 0.65\,\mas$.
This somewhat relaxes the above argument, but still requires
$\pi_\rel\ga 0.06\,\mas$, which puts the lens in the disk.

\section{{Physical Parameter Estimates}
\label{sec:phys}}

Because the microlens parallax $\bpi_\e$ is not measured, one cannot
directly infer the lens mass $M=\theta_\e/\kappa\pi_\e$ and lens-source
relative parallax $\pi_\rel = \theta_\e\pi_\e$ from the microlensing data.  
We therefore conduct a Bayesian analysis by incorporating priors from 
a Galactic model.  In fact, the situation is somewhat more complicated
than usual because the measurement error of the normalized source radius
$\rho$ is large and its distribution is asymmetric.
Moreover, $\rho$ is correlated with the planet-host mass ratio $q$.
Thus, in sharp contrast to the usual case, the posterior estimate
of this (seemingly) pure microlensing-light-curve parameter $q$ is
actually affected by the Galactic priors.

The fundamental features of the Galactic model and the Bayesian procedures
are the same\footnote{We note that \citet{ob171522} do not specify
their upper-mass cutoff for the mass function. For completeness, 
we note that this cutoff is $63\,M_\odot$.} as those presented in \citet{ob171522}. 
Here, we focus on describing what is different for the present case.

As usual, we incorporate constraints of both $t_\e$ and
$\theta_\e$ determined from the light-curve modeling and CMD. 
For this, we weight the simulated events by 
$W_{t_\e}=\exp[-(t_\e -t_{\e.\rm best})^2/2\sigma_{t_\e}^2]$,
where $t_\e$ is the Einstein timescale of the simulated event and 
$(t_{\e.\rm best},\sigma_{t_\e})$ are the values from Table~\ref{tab:ulens}.  

\begin{figure}
\plotone{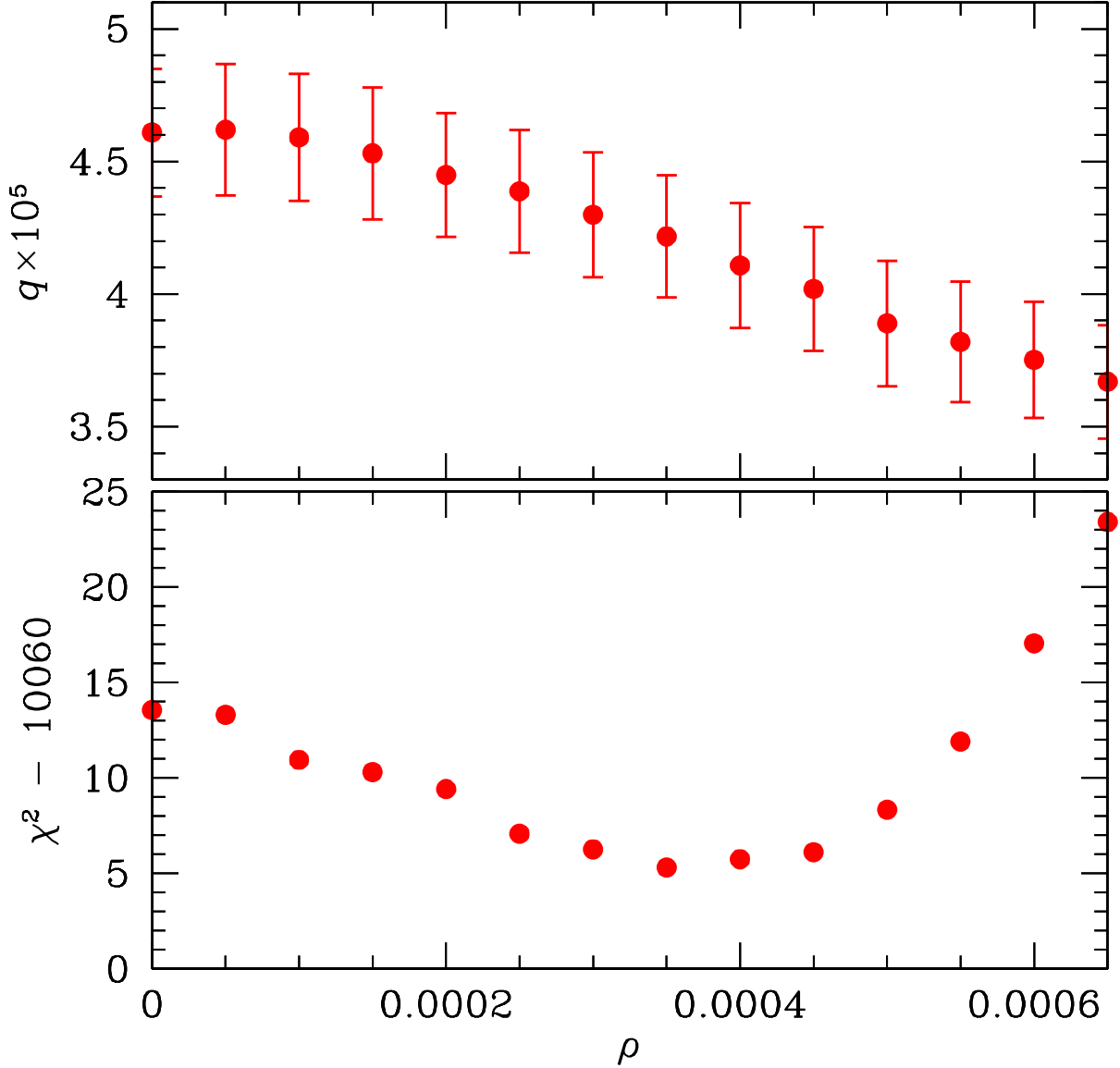}
\caption{Lower panel: minimum values of $\chi^2$ for a series of MCMC
runs with the normalized source size $\rho$ held fixed at the
indicated values, each with 10,000 accepted elements on the chain.
The minimum is overall relatively broad but high values (corresponding
to small Einstein radii $\theta_\e=\theta_*/\rho$) are strongly ruled out.
Upper panel: means and standard deviations of the planet-star mass ratio
$q$ for each of the MCMC runs carried out to make the bottom panel.
Note that $q$ is correlated with $\rho$ over the broad $\chi^2$
minimum of the latter.  This correlation is taken into account in the
Bayesian analysis (Section~\ref{sec:phys}).
}
\label{fig:grid}
\end{figure}

We incorporate the $\theta_{\rm E}$ constraint in a different way 
from that of the $t_{\rm E}$ constraint. 
We first evaluate a function $\Delta\chi^2(\rho)$ by running MCMCs 
for a series of fixed values of $\rho$. See Figure~\ref{fig:grid}. 
Then, for each simulated event, we evaluate $\theta_\e =\sqrt{\kappa M \pi_\rel}$ 
from the values of $M$, $D_{\rm L}$, and $D_{\rm S}$ of the individual simulated events. 
In this process, we draw $\theta_{*}$ from a Gaussian distribution 
with the mean and standard deviation presented in Equation~(\ref{eqn:thetastar}). 
We then evaluate $\rho=\theta_*/\theta_\e$ and finally determine the weight 
by $W_{\theta_\e} = \exp[-\Delta\chi^2(\rho)/2]$.

The Bayesian analysis then carried out following a usual procedure, 
but with one major exception. In Figure~\ref{fig:grid}, we show the best-fit
value of $q$ and its $1\,\sigma$ error bar for each fixed-$\rho$ MCMC, 
that is carried out following the procedure described above, to evaluate $\Delta\chi^2(\rho)$. 
Over the region of principal interest, $3\la \rho/10^{-4} \la 5$, this 
shows a logarithmic (power-law) gradient $d\log q/d\log \rho \simeq -0.2$. 
This gradient is present because the models account for the observed 
duration of the ``bump'' by the combination of $q$ and $\rho$: for 
relatively large $q$, this duration is explained by the width of the ridge, 
but with decreasing $q$ it must be increasingly explained by larger 
$\rho$. For most events, such a gradient would play little role because 
$\rho$ would be well determined. However, in the present case, the error 
in $\rho$ is large, so that the error in $q$ induced by this correlation 
is comparable to the scatter in $q$ at fixed $\rho$. Therefore, to 
properly evaluate the {\it posterior} value of $q$, we should incorporate 
the correlation into the Bayesian analysis.

We do so by evaluating $q$ separately for each simulated event in the
MCMC.  We first find $\rho$ (as above). We then find $q(\rho)$ and
$\sigma_q(\rho)$ from a table that corresponds to Figure~\ref{fig:grid}, 
and then draw $q$ randomly from a Gaussian described by these two values. 
This value of $q$ is then used both to evaluate the planet mass for the 
simulated event $m_p = qM_{\rm host}$ and to find the posterior distribution 
of $q$ itself.

The results of the Bayesian analysis are shown in Table~\ref{tab:phys} 
and Figure~\ref{fig:bayes}. 
Although the lens is believed to lie in the disk from the estimated $\theta_{\rm E}$, 
we formally consider both bulge and disk lenses in the Bayesian analysis. 
As expected, we find that the chance that the lens is in the bulge is
very low, $< 1\%$.

\subsection{{Does the Lens Account for the Blended Light?}
\label{sec:blend}}

The best-fit position of the blended light
$[(V-I),I]_b = (1.68,19.84)$ lies at a position with offsets 
$\Delta[(V-I),I]_b =[(V-I),I]_b -[(V-I),I]_{\rm cl} = (-0.75,3.80)$
from the red-clump-centroid on the CMD. 
We discuss the error bars on this position below.

The $I$-band brightness offset $\Delta I_b=3.80$ is consistent with
the lens mass and distance ranges derived from the Bayesian analysis
above, as summarized in Figure~\ref{fig:bayes}. 
More specifically, assuming the lens distance as the median of the
posterior distance distribution, at $D_{\rm L} = 3.3\,\kpc$, the lens would be 
at about $z\simeq -0.1\,\kpc$ below the galactic plane, and so plausibly
behind about half the dust toward the bulge, toward which 
$[E(V-I),A_I]=(1.28,1.50)$ \citep{nataf13}. This would imply an
absolute magnitude of the lens of 
$M_I = (I_{\rm cl,0}+\Delta{I})-A_{I}/2-[5{\rm log}\,(3300)-5] = 14.44 + 3.80 - 1.50/2 - 12.59 \simeq 4.9$. 
This corresponds roughly to an $M=0.85\,M_\odot$ star, which is quite
compatible with the Bayesian host-mass estimation. Such a star
would have (depending on its metallicity) roughly $(V-I)_0\sim 0.85$,
and therefore would be $\Delta(V-I) = 0.85-1.06 - 1.28/2\simeq -0.85$ 
mag bluer than the clump. Given the measurement errors (which we estimate 
just below), this is also quite compatible with
the ``observed'' offset: $-0.75$ mag bluer than the clump.

The baseline object is quite faint in the $V$ band, 
$V_{\rm base}=21.43$.  It is therefore barely detected, and hence
the error in magnitudes is large.  The contribution from the source
to this baseline light is known quite precisely, so from a conceptual
point of view, we should consider the errors in the point-spread-function (PSF) 
modeling of the remaining blended light $V_b= 21.52$.
The error in this estimate is comprised
of three distinct components: Poisson errors from finite photon
statistics; systematic errors from the PSF
modeling photometry program, operating under the assumption that 
the background is smooth; and statistical errors due to the fact
that a mottled distribution of unresolved stars contributes significantly
to the background.

Based on photon statistics, we estimate the systematic error as 0.19 mag. 
Because the PSF photometry program is relatively complex, it is difficult 
to reliably estimate the systematic errors. Hence, we ignore these errors 
for the moment. To estimate the statistical errors due to the mottled background,
we apply the approach of \citet{ob180532}, i.e., modeling the 
\citet{holtzman98} luminosity function adjusted for the local 
surface brightness of the bulge and the local extinction. In fact, 
the mottled background results in correlated errors between the 
$I$ and $V$ band measurements because there can be an excess or a 
``hole'' in this background of stars that are predominantly redder than 
the apparent blend star.  Hence, an excess at the location of the event 
would cause the apparent blend to appear brighter and redder, while a ``hole'' 
in the background would make it appear fainter and bluer.

Based on the surface density of clump stars (\citealt{nataf13}; D. Nataf, 2019,
private communication), we find a normalization factor of 2.41 relative
to Baade's Window. We adopt $[E(V-I),A_I]=(1.28,1.50)$ from \citet{nataf13},
and then evaluate the statistical errors using a $1.5^{\prime\prime}$ FWHM seeing disk. 
We find that (from this effect alone), there is a 16\% probability that the blend 
appears at least 0.20 mag bluer than it is, as well as
a 16\% probability that the blend appears at least 0.28 mag 
redder than it is. Hence, combining the two effects (systematic and statistical errors), 
the color error is at least $\pm 0.28$ mag.

We should also ask how well the source is astrometrically aligned with the
baseline object.  We measure the position of the source 
and baseline object relative to the KMTC42 
template and find in (west, north) $0.4^{\prime\prime}$ pixel coordinates, 
that the baseline object 
lies $(0.32,-0.12)$ pixels, or 
$(0.13^{\prime\prime},-0.05^{\prime\prime})$, i.e., west and south of the source.
Because the source position is derived from difference images (which
removes both the resolved and unresolved backgrounds) and because the
source is highly magnified in these images, the errors in the source
position are negligible relative to the errors in the baseline object.
Hence we ignore them.

The astrometric errors originate from the same three types of the photometric errors. 
The astrometry is done in the $I$-band, so we evaluate these errors in this band. 
We again only evaluate the errors of the first and third types. 
We estimate the fractional astrometric error 
(relative to the Gaussian half width $\sigma={\rm FWHM}/\sqrt{\ln 256}$) 
as that of the fractional photometric error, $(\ln 10/2.5)\sigma_I$.
That is, 
$\sigma_{\rm ast} = 0.39\sigma_I {\rm FWHM} \rightarrow 0.6^{\prime\prime}\sigma_I$.
We find that the photon-error contribution to $\sigma_I$ is 0.14 mag.

Next, we consider the error due to the mottled background. If we
evaluate this without any constraint, we find $\sigma_I= 0.6$. 
However, if we restrict to cases where the background produces a ``hole'', 
then $\sigma_I = 0.3$. Combined, these two
sources of error imply $\sigma_{\rm ast} = 0.37^{\prime\prime}$ and
$\sigma_{\rm ast} = 0.20^{\prime\prime}$ for the two cases.  This is 
larger than the observed offset. Hence, the measured astrometric offset 
is not inconsistent with the hypothesis that the blend is the lens. 
However, we note that this hypothesis can be established by 
future high-resolution observations.

\section{{Discussion}
\label{sec:discuss}}

\subsection{{Very Low Mass-Ratio Planets}
\label{sec:vlmq}}

It is found that the KMT-2019-BLG-0842Lb is a ``cold'' planet located 
beyond the snow line of its host, and the planet/host mass ratio is $\mathbf{q=(4.09\pm 0.27) \times 10^{-5}}$, 
which is similar to the ratio of Uranus/Sun in the Solar system. The discovery of the planetary system, 
together with similar systems previously discovered, provides an evidence that 
such planets are not rare. Nevertheless, further discoveries will be necessary 
to estimate the frequency and characterize the distribution of such planets.


As noted by \citet{ob180532}, the detection pace of 
low-mass-ratio microlensing planets with $q\leq 1\times 10^{-4}$ 
has been accelerating since 2015, when KMTNet commenced.
The 11 such planets (including KMT-2018-BLG-0029Lb and KMT-2019-BLG-0842Lb)
occurred in (2005, 2005, 2007, 2009, 2013, 2015, 2016, 2017, 2018, 2018, 2019).
That is, five during the 10 seasons before 2015 and six during the five
seasons after. Moreover, the data from the 2019 season 
have not yet been fully analyzed. If indeed sub-Uranus/Sun planets are common, 
we can expect more detections based on the trend of increasing rate.

In this context, it is worthwhile to ask what features of low mass-ratio planetary events 
allow them to be detected. A related issue is how the planet detection strategy should be 
adjusted to maximize the detection rate.


A notable feature of KMT-2019-BLG-0842Lb is that the planetary signal 
appeared in a high-magnification event $(u_0=0.0066;\ A_\max=150)$, and 
at very low angle $\alpha=0.146$ $(8.5^\circ)$
between the source trajectory and the 
binary axis.  That is, the magnification of the underlying 1L1S event
at the time of the planetary anomaly was modest: 
$A_{\rm anom} \simeq \sin(\alpha)/u_0 = 22$.  This means that
the source would have passed over the same part of the caustic
structure if the event had had $A_\max=22\ (u_0=0.045)$ and $\alpha=90^\circ$.
This is similar to the situations for OGLE-2016-BLG-1195, which had
$|u_0|=0.053$,  $|\alpha| = 55^\circ$, and a slightly higher mass ratio, 
and for
KMT-2018-BLG-0029, which had $|u_0|=0.027$,  $|\alpha|= 88^\circ$, and a 
substantially lower mass ratio.
Thus, at first sight, it seems
that such moderate-magnification events provide as fertile ground
to hunt for low-mass planets as high-magnification events (as advocated by
\citealt{abe13}).  In fact,
however, KMT-2019-BLG-0842Lb would have been discovered for values of
$\alpha$ covering most of the unit circle at its actual $u_0$, whereas
at $u_0=0.045$ the sensitivity would have been restricted to a 
relatively narrow range of angles.

However, the main feature to be noted is that the low source-trajectory 
angle caused the anomaly to be longer by factor $\cot\alpha \simeq 7$ 
relative to an orthogonal transit of the caustic structure. 
That is, in an orthogonal crossing, the full duration would have been 
about two hours rather than 15 hours. This is just twice 
the source-diameter self-crossing time, $2t_*\equiv 2\rho t_\e = 0.9\,{\rm hr}$. 
Such a short anomaly would have been detected in the actual observations 
(provided that it did not fall in a gap) because the event was in 
a high-cadence KMT field, with $\Gamma = 4$--$6 \,{\rm hr}^{-1}$. 
But if it had been in a field with a cadence $\Gamma = 0.75$--$1 \,{\rm hr}^{-1}$, 
then a two-hour anomaly would have been missed, while 
a 15-hour anomaly (as in this case) can be readily detected.

These points are in some sense moot because the survey strategy is basically set. 
However, these anomalies are not necessarily noticed in the mode by which
events are currently vetted, i.e., either manual or machine review
of pipeline data, with optimized photometry only for those events that ``look interesting''. 
The case of KMT-2019-BLG-0842 demonstrates that very low-$q$ planets 
can give rise to long-lived ($\ga 10\,$hr) low-amplitude ($\la 0.1\,$mag)
``bumps'' several mag below the peak of high-magnification events. 
Such planetary signals can be missed from the present search process. 
Hence, it would seem to be prudent to make optimized photometry of all high-mag events.

Moreover, there is a separate question as to how to marshal follow-up 
observations in order to probe to the lowest mass-ratio planets possible, 
i.e., planets of substantially smaller $q$ than those that have been 
detected to date.  At present, the rule of thumb is to intensively 
monitor high-magnification events within the FWHM around $t_0$ because 
this region contains ``most'' of the sensitivity to planets. 
Specifically, ``most'' translates to a fraction $\sim (\sec^{-1} 2)/(\pi/2)= 2/3$.
However, planets like KMT-2019-BLG-0842Lb lie in the other (remaining) 1/3 of the circle.
Although a minority, these planets can give rise to long lived perturbations
even when the planet has low mass ratio $q$, which can be of exceptional
interest. Thus, particularly in very high-magnification events,
for which $t_\eff$ is short, follow up observations during $t_\eff$ 
are less expensive to carry out. Therefore, follow-up observations 
should be more aggressively pursued for very high-magnification events.

\subsection{{Future High-Resolution Imaging and Spectroscopy}
\label{sec:future}}

As discussed in Section~\ref{sec:blend}, both the photometry and astrometry 
indicates the possibility that most of the blended light is generated 
by the lens. This possibility can be tested with high resolution
imaging, either with {\it Hubble Space Telescope (HST)} or ground-based
adaptive optics (AO) mounted on very large ground-based telescopes. Based on 
the Bayesian estimates together with the star catalog of \citet{pecaut13}, 
we estimate the dereddened lens magnitude in the $V$, $I$, and $H$ band as 
$(V, I, H)_{\rm L,0} = (18.78_{-1.90}^{+4.95}, 17.86_{-1.62}^{+3.45}, 16.75_{-1.19}^{+2.63})$. 
For the source, we estimate the dereddened magnitude as 
$(V, I, H)_{\rm S,0} = (21.14\pm0.07, 20.13\pm0.04, 18.85\pm0.07)$ based on our CMD analysis. 
These imply that the blended light is $\sim2$ magnitudes brighter than the source in the $I$ band.  
Therefore, if {\it HST} $I$-band images show that this blended light is 
closely aligned with the source position, then this will provide
strong evidence that the blended light is due to the lens. 
In principle, such aligned light may originate from a stellar companion to either the source
or the lens. However, if the alignment is very close (10--20 mas), 
then this will rule out the possibility of the lens companion 
because such a companion would have generated significant deviations over the 
well-covered peak of the event. The possibility of the source companion can 
ultimately be ruled out by re-imaging the field when the source and blend 
are separated far enough to measure their relative proper motion. 
Because the source and blend have substantially different colors, such
a measurement does not require the two stars to be separately resolved,
but can be carried out via a measurement of the astrometric
offset between their combined light in different bands \citep{ob03235b}.

\begin{figure}
\plotone{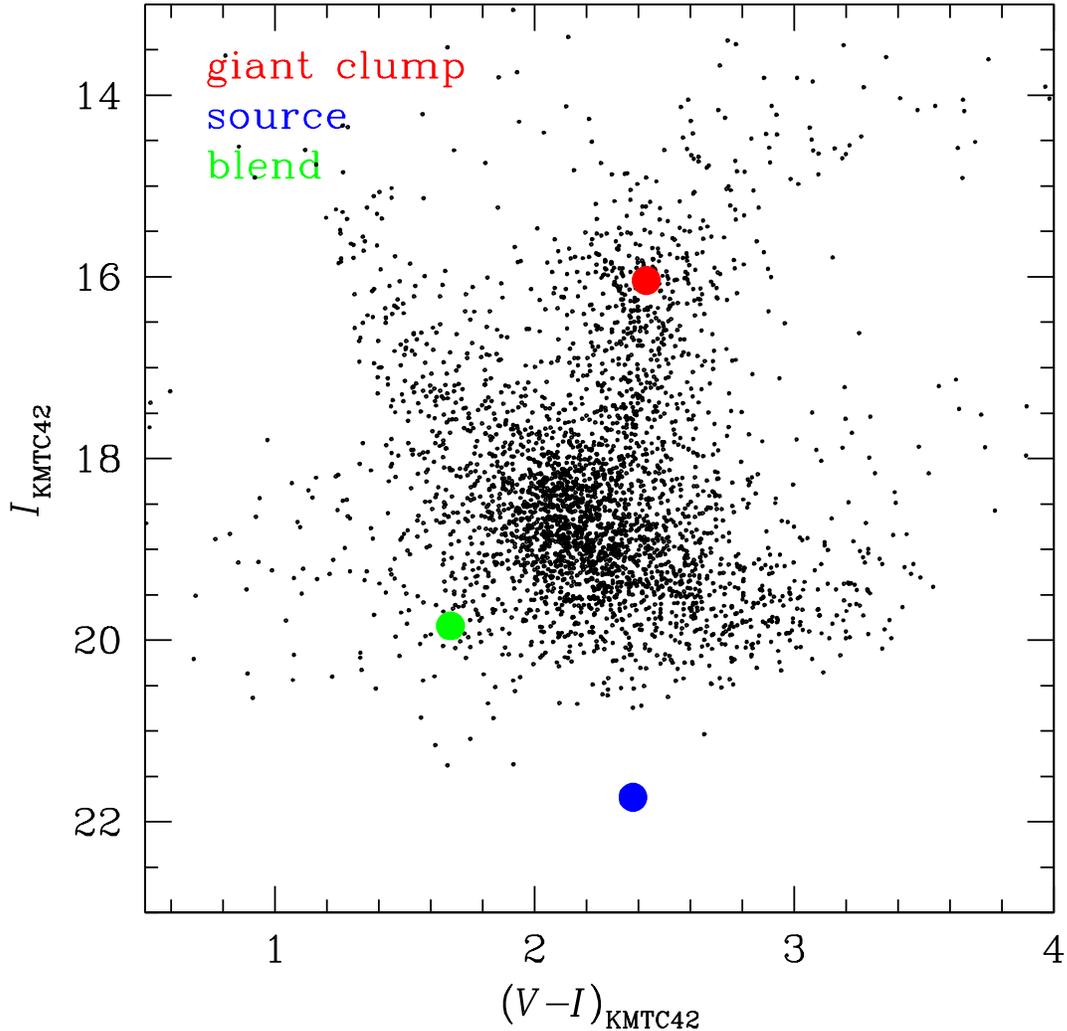}
\caption{Color-magnitude diagram (CMD) for field stars in a $2^\prime$
square centered on KMT-2019-BLG-0842 derived from pyDIA reductions of
KMTC42 data.  The positions of the source (blue), the blend (green),
and the centroid of the red clump (red) are indicated.  The error bars
for these determinations are discussed in the text.  In particular,
the blend color is relatively poorly determined.  By cross-matching to
the calibrated OGLE-III catalog \citep{ogleiii1,ogleiii2}, we find
$I_{\rm calib} = I_{\rm KMTC42C} -0.085$,
$(V-I)_{\rm calib} = (V-I)_{\rm KMTC42C} -0.140$.
}
\label{fig:cmd}
\end{figure}

Ground-based AO astrometry will also be useful to check whether the blended 
light is aligned with the lens. The excess flux from the blend will be 
somewhat more difficult to be measured in this case because such measurements 
will be conducted in near-infrared bands (e.g., $H$ band),
for which there is no direct measurement of the source flux. 
Nevertheless, based on $VIH$ color-color relations of nearby field stars, 
derived from the KMTC42 CMD (Figure~\ref{fig:cmd}) and the future AO measurements
aligned to standard catalogs, it should be possible to predict the $H$-band source 
flux with reasonably good precision. Even though the blended light is much bluer than 
the source, it should still be substantially brighter in the $H$ band than the
source (because it is 2 mag brighter in the $I$ band). 
Thus, ground-based AO measurements should be feasible.

The Bayesian posterior estimate of the lens-source relative proper
motion is $\mu_\rel=8.4^{+1.7}_{-1.2}\,\masyr$ (Table~\ref{tab:phys}).
Hence, to obtain the first measurement when the lens and source are
still closely aligned, the observations should be made in 2020, if
possible.

If the blended light proves to be aligned with the source, then it should
be possible to spectroscopically classify the blend \citep{ob180740}. 
Such an observation may also plausibly measure the radial velocity (RV) offset 
between the source and the blend. If so, this will provide a second 
(and earlier) method to rule out the blend as being a companion to the source.
Note that, to avoid generating a noticeable bump on the light curve,
the blend (as companion) must have a projected separation $\ga 8\,\au$, 
and so an RV offset of $\la 15\,\kms$ relative to the source.

\begin{figure}
\plotone{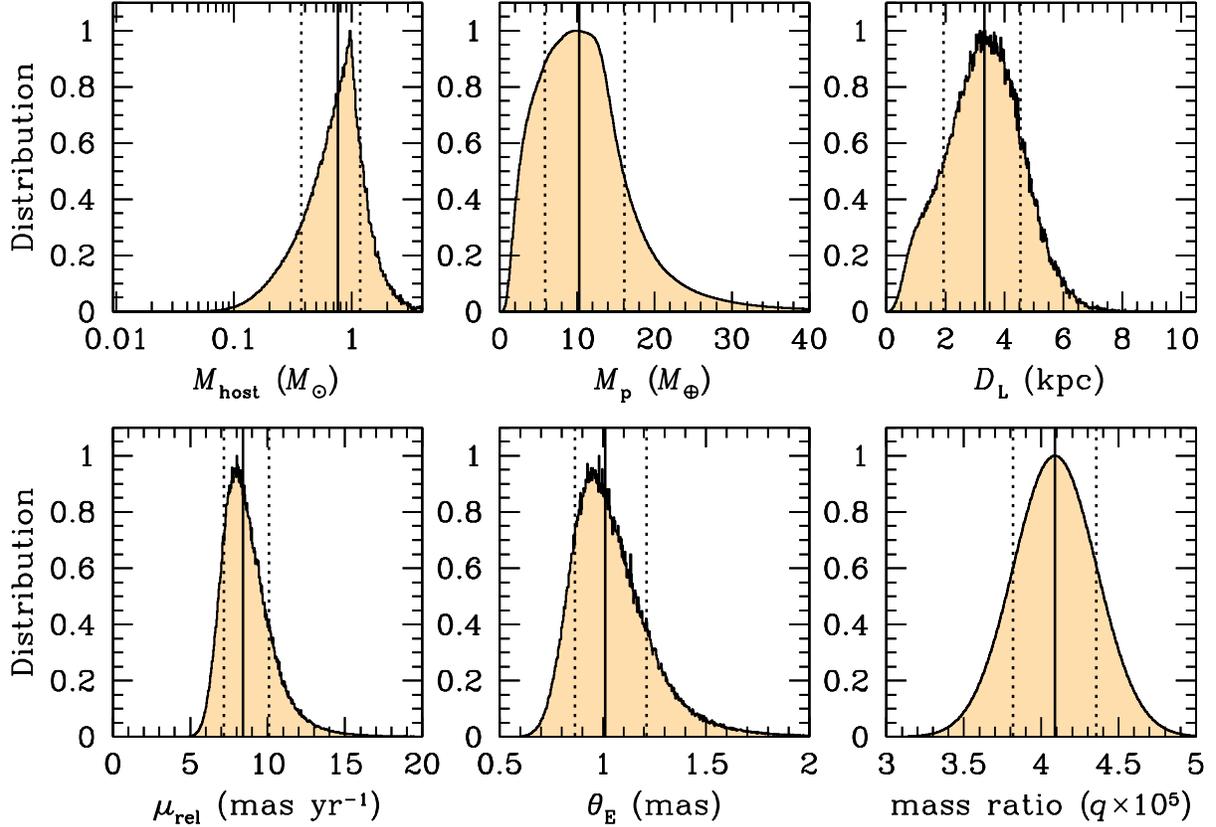}
\caption{Bayesian posteriors for three physical parameters 
(host mass, planet mass, and system distance), along with three
parameters that would normally be derived directly from the light
curve and CMD (lens-source relative proper motion $\mu_\rel$,
Einstein radius $\theta_\e$, and
planet-star mass ratio $q$).  Bayesian posteriors
are required for these three because $\rho$ is relatively poorly determined
and with significantly asymmetric errors.  In particular, $q$ is correlated
with $\rho$ (see Figure~\ref{fig:grid}).
}
\label{fig:bayes}
\end{figure}

If the blend proves not to be the lens (or, more precisely, not
to be dominated by the lens light), then it should still be possible to
characterize the lens with high-resolution followup.  Depending on
the flux and/or color differences between the source and the lens,
this may be achieved even before the images are fully separated, e.g.,
by measuring the astrometric offset between the combined images in different
bands \citep{ob03235b} or by measuring the elongation of the combined
image \citep{ob120950}.  Even if the flux ratio is extreme, causing these
combined-image techniques to fail, the lens can still be
separately imaged when it has moved $\sim 60\,$mas \citep{ob05169bat}
from the source, i.e., roughly by 2027.  From Figure~\ref{fig:bayes}, the lens
mass is almost certainly above the hydrogen-burning limit, so
this method is almost guaranteed to work if all others fail.



\acknowledgments 
W.Z. acknowledges support by the National Science Foundation of China (Grant No. 11821303 and 11761131004)
Work by AG was supported by AST-1516842 from the US NSF and by JPL grant 1500811.
AG received support from the European  Research  Council  under  the  European  Union’s Seventh Framework Programme (FP 7) ERC Grant Agreement n. [321035]
This research has made use of the KMTNet system operated by the Korea
Astronomy and Space Science Institute (KASI) and the data were obtained at
three host sites of CTIO in Chile, SAAO in South Africa, and SSO in
Australia.
Work by CH was supported by the grants of National Research Foundation of Korea (2017R1A4A1015178 and 2020R1A4A2002885).
%
The OGLE project has received funding from the National Science Centre,
Poland, grant MAESTRO 2014/14/A/ST9/00121 to AU.
The MOA project is supported by JSPS KAKENHI Grant Number JSPS24253004, JSPS26247023, JSPS23340064, JSPS15H00781, and JP16H06287.
Work by CR was supported by an appointment to the NASA Postdoctoral Program
at the Goddard Space Flight Center, administered by USRA through a contract
with NASA.

\begin{deluxetable}{lrr}
\tablecaption{Binary Lens Models}
\tablewidth{0pt}
\tablehead{
\multicolumn{1}{l}{Parameters} &
\multicolumn{1}{c}{Close} &
\multicolumn{1}{c}{Wide}
}
\startdata
$\chi^2$/dof            &    10065.4/10332       &     10241.5/10332       \\
$t_{0}$ (${\rm HJD'}$)  & 8625.914 $\pm$ 0.010   &  8625.914 $\pm$ 0.011   \\
$u_{0}$ ($10^{-3}$)     &   6.565  $\pm$ 0.202   &     5.393 $\pm$ 0.183   \\
$t_{\rm E}$ (days)      &  43.859  $\pm$ 1.287   &    52.915 $\pm$ 1.708   \\
$s$                     &   0.983  $\pm$ 0.013   &     1.063 $\pm$ 0.013   \\
$q$ ($10^{-5}$)         &   4.079  $\pm$ 0.289   &     3.901 $\pm$ 0.249   \\
$\alpha$ (rad)          &   0.146  $\pm$ 0.067   &     0.144 $\pm$ 0.072   \\
$\rho$ ($10^{-4}$) &   4.246  $\pm$ 0.966   &     $0.282_{-0.090}^{+0.913}$ \\                  
$f_{{\rm s}}$           &   0.037                &     0.030              \\
$f_{{\rm b}}$           &   0.179                &     0.185 
\enddata  
\vspace{0.05cm}
\tablecomments{
${\rm HJD}'= {\rm HJD}-2450000~{\rm days}$
}
\label{tab:ulens}
\end{deluxetable}

\begin{deluxetable}{lr}
\tablecaption{Binary Source Model}
\tablewidth{0pt}
\tablehead{
\multicolumn{1}{l}{Parameters} &
\multicolumn{1}{c}{1L2S}
}

\startdata
$\chi^2$/dof                     &     10560.1/10031     \\
$t_{0,1}$ (HJD$'$)               &  8625.907$\pm$0.013   \\
$u_{0,1}$ $(10^{-3})$            &     5.620$\pm$0.213   \\
$t_{0,2}$ (HJD$'$)               &  8627.886$\pm$0.032   \\
$u_{0,2}$ $(10^{-3})$            &     0.005$\pm$0.247  \\
$t_{\rm E}$ (days)               &    48.657$\pm$1.710   \\
$\rho_{1}$ $(10^{-4})$         &     $7.6_{-1.5}^{+16.0}$  \\
$\rho_{2}$ $(10^{-4})$         &     32.3$\pm$2.2   \\
$q_{F,I}$ $(10^{-2})$            &     1.291$\pm$0.043   \\
$f_{{\rm s}}$                    &     0.033             \\
$f_{{\rm b}}$                    &     0.183 
\enddata  
\vspace{0.05cm}
\label{tab:1L2S}
\end{deluxetable}

\begin{deluxetable}{lr}
\tablecaption{Lens properties}
\tablewidth{0pt}
\tablehead{
\multicolumn{1}{l}{Parameters} &
\multicolumn{1}{c}{Values} 
}
\startdata
$q$ $(10^{-5})$                                      &   $4.09_{-0.27}^{+0.26}$   \\ 
$\theta_{\rm E}$ (mas)                               &   $1.01_{-0.15}^{+0.20}$   \\ 
$\mu_{\rm rel}$ $({\rm mas}\,{\rm yr}^{-1})$         &   $8.42_{-1.24}^{+1.68}$   \\ 
$M_{{\rm host}}$   $(M_{\odot})$                     &   $0.76_{-0.39}^{+0.42}$  \\ 
$M_{{\rm planet}}$ $(M_{\oplus})$                    &   $10.28_{-5.28}^{+5.84}$ \\ 
$a_{\bot}$ (AU)                                      &   $3.31_{-1.47}^{+1.38}$  \\
$D_{\rm L}$ (kpc)                                    &   $3.32_{-1.39}^{+1.22}$  
\enddata 
\vspace{0.05cm}
\label{tab:phys}
\end{deluxetable}

\end{document}